\documentclass{aa}
\usepackage{graphicx}
\newcommand{\kms}{\mbox{km\,s$^{-1}$\,}}
\newcommand{\masyr}{\mbox{mas\,yr$^{-1}$\,}}
\begin{document}
\title{NGC 6994 - clearly not a physical stellar ensemble
\thanks{Based on observations made with the spectrograph Elodie
           on the 1.93m telescope at Observatoire de Haute Provence}}


\author{M. Odenkirchen\inst{1}
          \and
        C. Soubiran\inst{2}}

\offprints{M. Odenkirchen}

\institute{Max-Planck-Institut f\"ur Astronomie, 
           K\"onigstuhl 17, D-69117 Heidelberg, Germany\\
           \email{odenkirchen@mpia-hd.mg.de}
    \and
       Observatoire de Bordeaux, UMR 5804, BP 89, F-33270 Floirac, France\\
       \email{soubiran@observ.u-bordeaux.fr}}

\date{Received 24 August 2001; accepted 30 November 2001}

\abstract{
The sparse stellar ensemble NGC 6994 = M\,73 has recently been discussed as 
a possible remnant of an old open cluster.
In order to solve the controversy on the nature of this object we have
taken high-resolution spectra of the six brightest stars within $6'$
angular distance of its nominal position.
These stars are the only obvious member candidates for an eventual cluster 
or cluster remnant since fainter stars do not show any significant 
concentration in the plane of the sky.
The radial velocities, atmospheric parameters, and absolute magnitudes 
derived from the spectra reveal that the six candidates do not share 
the same kinematics and lie at different distances from the Sun.
The proper motions provided by the Tycho-2 catalogue show that there is also 
a large spread in the tangential motions of these stars, in agreement
with the spectroscopic results.
This leads to the conclusion that the few bright stars that constitute 
NGC\,6994 are not a physical system (cluster, cluster remnant, or group). 
They must instead be understood as a projective chance alignment of 
physically unrelated field stars. 
}
\authorrunning{Odenkirchen \& Soubiran}
\titlerunning{NGC\,6994}
\maketitle


\section{Introduction}
There is convincing evidence, from numerical simulations 
(e.g. Terlevich 1987, de la Fuente Marcos 1997) as well as from the 
observed distribution of cluster ages (Wielen 1971, Janes \& Adler 1982, 
Lyng{\aa} 1982), that most of the open star clusters in the Milky Way 
undergo rapid dynamical evolution and decay on a time scale of 200 --
500~Myr.
The dynamical evolution of an open cluster is driven (1) by mass-loss 
from stellar evolution of its most massive members, (2) by 
internal energy repartition in two-body interactions and encounters 
with binaries, and (3) by external perturbations due to the tidal
forces exerted by the mean galactic field and by local inhomogeneities.
In the final stage of its evolution an open cluster is expected to 
reduce to a small number of relatively massive stars, preferentially 
binaries, which have sunk to the cluster's core while the less massive 
members have progressively escaped into the galactic field 
(de la Fuente Marcos 1997).  
Since the Galactic disk contains several thousand open clusters and since
the typical lifetimes of these clusters are short with respect 
to the age of the disk, it seems likely that such cluster remnants 
exist in large numbers. 
It would be important to identify and investigate these remnants in 
order to check and constrain the numerical models of cluster evolution. 
However such small groups of stars are in general difficult to detect 
because they are seen in projection onto the field of foreground and 
background stars which dominate the stellar surface density, in particular 
in regions close to the galactic plane.\\

An interesting candidate for a very sparse cluster or cluster remnant is 
the object NGC\,6994 = M\,73 ($\alpha,\delta = 20^h58\fm 9, -12^\circ38'$ 
(J2000.0); $l,b = 35\fdg 7, -34\fdg 0$) which is located at fairly high 
galactic latitude.
It was first described by Messier (1784) as ``a cluster of three or four 
small stars which resemble a nebula at first sight''. 
Later observers specified that it consists of four stars of magnitude 
10 - 13 in an oblique triangle and that the two brightest of these stars  
are BD$-$13$^\circ$\,5808 and BD$-$13$^\circ$\,5809 (Holetschek 1907, 
Reinmuth 1926).  
The object has since been listed in a number of catalogs of open clusters 
(e.g.\ Collinder 1931, Alter et al.\ 1958, Lyng{\aa} 1987).
Collinder (1931) gave the diameter of the cluster as $2'$ to $3'$, 
but somehow presumed that it might contain about 200 members and perhaps 
be a globular cluster. 
Ruprecht (1966) on the other hand classified it as a barely conspicuous 
poor cluster with stars of almost equal brightness and marked it as a 
doubtful case. 
Bassino et al.\ (2000) and Carraro (2000) recently undertook the first deep 
photometric studies of NGC\,6994. These studies ended up with contradictory
conclusions on the physical reality of the cluster. 
Same as Bassino et al. (2000),  Bica et al.\ (2001) recently proposed 
NGC\,6994 as a possible open cluster remnant. \\
 
In order to obtain a clear answer on the nature of NGC\,6994, we decided to 
take high-resolution spectra of the few obvious bright member candidates 
and to derive their precise radial velocities and absolute magnitudes. 
In addition we also take into account the proper motions of the candidates
as an important kinematic diagnostic. Clearly, small groups of physically 
related stars that do not stand out strongly against the underlying field 
can best be recognized by finding out whether the supposed members lie at 
a common distance and share a common space motion.
In Section 2 of this paper we briefly review the different arguments on 
NGC\,6994 that have been given in previous studies.
In Section 3, we investigate the significance of NGC\,6994 from the 
viewpoints of stellar surface density in different magnitude intervals
and number statistics of bright stars.
In Section 4 we consider the information available from proper motions. 
Section 5 then describes our spectroscopic observations and the results
derived from them.
In Section 6 we summarize our results and conclusions.\\

\section{Previous studies on NGC\,6994}

Detailed investigations of the region of NGC\,6994 exist only since 
recently.
Bassino et al.\ (2000) have presented color-magnitude diagrams (CMDs) 
from CCD photometry in $BVRI$ down to $V$ = 21~mag for a field of diameter 
$9\farcm 6$. 
They argue that the distribution of stars in their CMDs provides evidence 
for the presence of an old open cluster.  
The cluster is supposed to have an angular diameter of at least $9'$, a 
distance of about 620\,pc and an age of 2 to 3 Gyr. The members of the 
cluster are however not directly obvious from the structure of the CMD, 
and a CMD for a nearby comparison field, which would in this case be 
essential, is not provided.  
Twenty-four stars are proposed as member candidates, in particular all 
stars brighter than $V = 14.5$ in the field. These candidates have been 
selected with an iterative process of isochrone fitting in the plane of 
$(B-V,V)$, but without the ability to take into account the 
color-magnitude distribution of field stars. In order to validate the 
result of this selection scheme the authors refer to predictions from a 
Galactic model by Reid \& Majewski (1993), showing that their cluster 
member candidates are located on a sequence that lies away from the loci 
of simulated field stars. 
A critical point in this approach is that their comparison with the model 
relies on only one simulated sample. 
Since the field size is very small a single simulation may involve large 
fluctuations and so the CMD of only one simulated sample may not be 
representative. Our own experience from twenty simulations for the field 
of NGC\,6994 obtained with the Besan\c{c}on model of Galactic stellar 
population synthesis (Robin \& Cr\'ez\'e 1986\footnote{The latest 
version of this model is available at 
http://www.obs-besancon.fr/modele/modele\_ang.html}) shows that 
there are indeed considerable variations from sample to sample and 
no evidence for a general disagreement between the simulated field CMD's 
and the observed CMD.\\

The second argument by which Bassino et al.\ support their assumption 
of the existence of a cluster is the shape of the radial stellar surface 
density profile, which they show in their Figure 11. Including all stars 
from magnitude 10 to 21 within their field of radius $4\farcm 8$, the 
profile shows a clear increase of the surface density within $1'$ or 
$1\farcm 5$ of the center and a rather flat distribution outside this 
range. Bassino et al.\ interpret this as a proof for the existence of a 
cluster with radius $\ge 4\farcm 5$. 
However, without evidence for a significantly lower surface density at 
radii beyond $4\farcm 5$ an alternative interpretation of this density 
profile is that it reveals merely the known concentration 
of four bright stars within about $1'$ of the center plus a constant 
background density of field stars and no cluster of fainter stars.
In order to verify the cluster hypothesis the stellar surface density 
profile needs to be investigated on a larger scale and in different 
magnitude ranges. (see Section 3 and Fig.\,2). 
\\ 

In a similar study, Carraro (2000) has obtained CCD photometry in $BVI$ 
down to $V = 22$~mag in a field of $9' \times 9'$. While the resulting 
color-magnitude diagrams look very much like those of Bassino et al., 
Carraro draws a completely different conclusion from his data. 
He claims to see a normal field star CMD without any distinct cluster 
feature. He thus declares NGC\,6994 to be a random enhancement of four 
field stars rather than an open cluster or cluster remnant.
On the other hand, he also states that 11 stars in his sample significantly 
emerge from the field. This is confusing since it appears to be an argument 
that is in favor of the existence of a cluster and thus in conflict with the 
former conclusion. Unfortunately, the essential comparison with the stellar 
content in at least one nearby control field is again missing. Hence there 
is no obvious way to recognize whether or not there is a significantly 
enhanced number of stars and whether the distribution of stars in the 
CMD is consistent or inconsistent with that of the field star population. 
Furthermore, Carraro's statement that the two brightest stars in the 
center of NGC 6994 belong to the solar neighborhood and have distances of 
42~pc and 135~pc does not hold (see Section 5). 
The quoted distances are spurious because they have been derived from the 
very coarse (and in this case meaningless) trigonometric parallaxes given  
by the Tycho catalog (median standard error of 30 - 50~mas in the magnitude 
range in question; see ESA 1997, p.142). 
In summary, neither of the two studies have provided clear indications of 
what NGC\,6994 really is and which stars belong to it. Thus further 
investigations were required to find the answer. \\

Bica et al.\ (2001) have selected NGC\,6994 and a number of other known
stellar groups as possible open cluster remnants on the basis of enhanced 
surface density with respect to the surrounding field. 
They point out that the number of stars counted down to $B=12.5$~mag in a 
field of $9' \times 9'$ around NGC\,6994 is five instead of only one that 
would be expected from the average counts in a larger surrounding area. 
This alone is however not an argument for the physical nature of this 
group since any distribution of stars goes along with local fluctuations 
so that enhancements of low numbers can turn up also randomly. 
It is therefore necessary to quantify the expected frequency of such 
fluctuations, as will be done in the next section.\\

\section{Spatial configuration, surface density and number statistics}

We begin our study with a reanalysis of the spatial distribution of stars 
in the field of NGC\,6994 and around it using data from the USNO-A2.0 
star catalogue (Monet et al. 1998; based on POSS-I) and the Tycho-2 
star catalogue (H{\o}g et al.\ 2000). 
Fig.~1 shows the distribution of stars in a field of $30' \times 30'$ 
around NGC\,6994 down to $R = 17.5$\,mag according to USNO-A2.0.
Fig.~4a shows the stars from Tycho-2 in a somewhat larger field 
of $54' \times 30'$. 
Both reveal that the field of NGC\,6994 is characterized by an unusual group 
of four bright stars, with magnitudes in the range $10.4 \le V_T \le 11.9$ 
and mutual angular separations between $0\farcm 4$ and $1\farcm 0$. 
The mean position of this group is regarded as the center of NGC\,6994. 
The two brightest of its members are likely to correspond to the pair 
BD$-$13$^\circ$ 5808/5809 although we note that the BD positions are not 
entirely consistent with the real stellar configuration.  
Apart from this group, there are two more stars of similar brightness 
($11.1 \le V_T \le 12.4$) about $4'$ and $5'$ to the south.
These two may also be suspected to belong to the NGC\,6994 complex. 
All other stars out to more than $6'$ angular distance from the central 
group are at least 1~mag fainter than the aforementioned ones. \\

In Fig.~2 we show the surface densities of stars of different magnitudes 
as a function of angular distance from the center of NGC\,6994 (based on 
USNO-A2.0).
The central group of bright stars produces a strong local enhancement 
in the surface density of stars with magnitudes $R \le\ 12.5$ (see Fig.~2a).  
Fainter stars however, i.e., in the magnitude ranges from 12.5 to 15.0 and 
from 15.0 to 17.5 in $R$ do not show signs of a spatial concentration 
towards NGC\,6994 and therefore do not provide evidence for the presence
of a cluster. Figs.~2b and 2c show that the surface densities 
of stars in these magnitude intervals do not increase towards the center, 
but rather agree with the level of the surface density of surrounding field 
stars within the statistically expected uncertainty. 
In the faintest magnitude range there is a certain overdensity in the 
distance range from $3'$ to $4'$, but too close to the limit of statistical 
noise to be considered as a proof of the existence of a cluster ($< 2\sigma$). 
In addition, we will show in Section 6 that the color-magnitude diagram for 
the field of NGC\,6994 does also not provide evidence for cluster stars 
with magnitudes beyond 12.5 since there is actually no sign of a 
characteristic cluster main sequence.
It therefore appears that the significance of NGC\,6994 as a stellar 
ensemble is based solely on the clustering of the few stars brighter than 
12.5\,mag and that these are the only obvious member candidates for a 
cluster or cluster remnant in this place. 
Fig.~3 and Table~1 provide an identification of these stars in terms of 
position and Tycho catalogue number (TYC). Table~1 also shows the 
Tycho-2 photometry and proper motions of these stars.
We remark that although stars fainter than 12.5\,mag do not indicate
the existence of a cluster, it could nevertheless be that a few of the 
fainter stars are physically related to the brighter stars provided that 
the latter indeed form a physical group. \\

One may expect that the clustering of the bright stars in the plane of the 
sky is by itself a proof of the physical nature of NGC\,6994. However, 
it turns out that this argument is not as strong as it appears.  
The statistics of objects in the Tycho-2 catalog shows that the mean 
surface density of stars with magnitude $10.0 \le V_T \le 12.0$ in the 
region of NGC\,6994 is $\Sigma \simeq 38$~stars/deg$^2$, including some 
allowance for the incompleteness of the Tycho-2 catalog towards the faint 
end of the above interval. For a random distribution of stars with 
mean surface density $\Sigma$, i.e., a distribution with number fluctuations 
that follows Poisson statistics, the probability of finding $n$ neighbors 
in a circular field of radius $\Theta$ around an arbitrarily selected star 
is given by: 

\begin{equation}
p(n) = \frac{1}{n!}(\Sigma\pi\Theta^2)^n \exp(-\Sigma\pi\Theta^2)
\end{equation}

Using $\Sigma = 38$~stars/deg$^2$ and $\Theta = 1'$, the probabilities 
for $n$=1,2,3 are $p(1)=3.0\times 10^{-2}$, $p(2)=4.5\times 10^{-4}$, and 
$p(3)=4.6\times 10^{-6}$. 
This means that the relative frequency for a random occurrence of a 
quadruple of stars like in NGC\,6994 is about $5\times 10^{-6}$. 
In other words, to find one such configuration randomly one needs a sample 
of about 170,000 stars or 4500 square degrees of sky with mean surface 
density of 38~stars/deg$^2$. 
Therefore, from a statistical point of view the occurrence of NGC\,6994 
as a random grouping of physically unrelated stars is expected to be a 
rare case, but not so extremely rare that this possibility can be totally
excluded.
Turning it the other way round, the number statistics does not necessarily 
require that this particular group must be formed by physically related stars.
Therefore, only proper motions and radial velocities can settle the question. 
\\

\section{Evidence from proper motions}

Important information on the kinematics of the bright stars in and around 
NGC\,6994 is provided by the proper motions available in the Tycho-2 
catalogue. 
The Tycho-2 proper motions are based on the comparison between contemporary 
mean positions derived from the recent Tycho observations onboard Hipparcos 
and early-epoch positions observed many decades ago (see H{\o}g et al.\ 2000 
and references therein). 
Due to the long time-baseline they (1) have rather high precision (in 
contrast to the Tycho parallaxes mentioned in Section 2) and (2) are not 
subject to influences from short period orbital motion that - in the case of 
binaries - might otherwise substantially affect the observed photocentric 
motion (see e.g.\ Odenkirchen \& Brosche 1999). 
In other words, the proper motion data from Tycho-2 directly indicate the 
long-term mean tangential motions of the stars. 

In Fig.~4c the proper motions of all Tycho-2 stars in the $30' \times 54'$ 
field around NGC\,6994 are shown in a vector point diagram.
The six cluster members candidates from Table 1 are marked and identified 
by their TYC number. 
Fig.~4c clearly demonstrates that these stars do not share a common mean 
tangential motion. The differences between the individual proper motion 
vectors are between 5 and 28 \masyr while the typical uncertainty of the 
proper motions is about 2~\masyr per component. 
Depending on the distances of the stars the observed differences in proper 
motion correspond to differences in tangential space velocity of 5~\kms to 
28~\kms ($d=200$~pc) or 20~\kms to 112~\kms ($d=800$~pc). 
Such velocity differences strongly exceed the velocity dispersion of an open 
cluster or a long-lived group and are more characteristic of the general 
velocity dispersion in the Galactic disk.
Comparing the proper motions of the candidate stars to those of other Tycho 
stars in the surrounding field it is indeed seen that the spread of proper 
motions among the candidates is in agreement with the typical spread in the 
local distribution of field star proper motions. 
In summary, we find that the tangential motions of the candidate stars do not 
support the idea of a physical relationship between them but are 
rather favorable to the alternative assumption that these stars are part of 
the local field population.

\section{Spectroscopic investigation}

For further assessment of the kinematic coherence or incoherence of 
the six principal member candidates of NGC\,6994 we recently obtained 
high-resolution spectra of using the echelle spectrograph Elodie on the 
1.93m telescope of the Observatoire de Haute Provence (Baranne et al. 1996).
All stars were observed at two epochs, in July 2000 and August 2001. 
The spectra cover the range from 390 nm to 680 nm at a resolving power of 
42\,000 and have signal-to-noise ratios of 5.4 to 16.4 at 
$\lambda \simeq 555$~nm. 
The spectra were processed and calibrated with dedicated on-line data 
reduction software available at the telescope. 
Radial velocities were derived by crosscorrelating the spectra with a digital
template mask for spectral type F or K. There is abundant experience with the 
instrument showing that the accuracy of the radial velocities obtained in 
this way is better than 1 \kms for the considered stars of type F, G and K, 
even at relatively low signal-to-noise ratios (Baranne et al.\ 1996, 
Delfosse et al.\ 1999).  
Estimates of the atmospheric parameters $T_{eff}$, $\log g$ and [Fe/H] and 
the absolute magnitude $M_V$ of the stars were determined with the 
special TGMET software (Katz et al. 1998) in the same way as described 
in our previous open cluster study (Soubiran et al.\ 2000). Briefly, 
this software compares the observed spectrum to a library of empirical 
reference spectra and calculates the stellar parameters by weighted averaging 
over the parameter values of the five to ten best-matching reference stars. 
      
The results of the reduction of the spectra are listed in Table 2.
It turns out that there are no substantial variations in the radial 
velocities and the other stellar parameters between the two epochs for 
any of the six targets. 
The largest difference in radial velocity that occurs when comparing 
the results of the observations taken 13 months apart is about 0.8~\kms
(in this case the spectral lines are broadened due to rotation), 
the other differences are $\le 0.3$~\kms. This assures that the velocity 
measurements are not significantly affected by orbital motion with a 
companion and hence directly represent the mean space velocities of the 
stars.  

The radial velocities of the six targets cover the range from 
$-$54~\kms to +35~\kms. We find that there is no agreement between any of 
the velocities at the level of the uncertainty of the measurements 
which is $\le 1$~\kms. 
The two most closely coinciding radial velocities have a difference of 
4.1~\kms (stars 0802 and 0565).
There are three stars (0594, 0802, 0565) with radial velocities 
falling into the interval from $-$31~\kms to $-$21~\kms. 
The velocities of the other three stars are more widely dispersed. 
The three stars with somewhat similar velocities may at first sight 
appear as a kinematic subgroup. However, the differences in radial velocity 
are actually too large to interpret these stars as a wide triple system or 
a long-lived unbound group. Also, the proper motions and distances of these 
stars (see below) are such that their tangential velocities must be even 
more deviant than their radial velocities. This is true in particular 
for stars 0565 and 0802 which have proper motions that differ by 28\,\masyr
and hence can definitely not be part of the same kinematic group.     

The values of the atmospheric parameters $\log g$ and $T_{eff}$ 
reveal that the sample of candidates contains three dwarfs, two red giants, 
and one subgiant. The absolute luminosities of the stars differ by up to 
3.9~mag in $V$. This contrasts with a much smaller spread in the observed 
apparent $V$ magnitude which is less than 2~mag.        
Comparing the spectroscopically determined absolute magnitude $M_V$
and the photometrically measured $V$ (standard $V$ derived from Tycho-2 
photometry via the transformation given in ESA 1997, p.63) one obtains an 
estimate of the individual distance modulus of each star. 
The distance moduli are found to spread over 3~mag, ranging from 7.2~mag 
to 10.2~mag (see Table 2). 
This indicates that the distances of the stars differ by a factor of 4. 
According to the reddening maps of Schlegel et al.\ (1998) the reddening at 
the position of NGC\,6994 is $E(B-V) \simeq 0.045$~mag and varies between 
0.038~mag and 0.065~mag in the $20'\times 20'$ arcmin region around it. 
We thus assume that the extinction is near $A_V = 0.15$~mag so that the 
unreddened distance moduli are only slightly smaller than the measured ones. 
The distances of the stars hence lie between 250~pc (0492) and 1~kpc (0796). 
Altogether we find that the distances and radial velocities of the six stars 
strongly contradict the hypothesis that they form a physical group and does 
not even provide support for the existence of a subgroup 
or a physical pair. 

The metallicities of the six stars range from zero to about $-$0.3 dex.
The spread in metallicity suggests that the sample as a whole is not 
a chemically coherent group. It is possible to define subgroups 
of stars with similar metallicity, but these subgroups do not consistently 
match with the distributions of the other parameters, i.e. distance, radial 
velocity and tangential motion, and thus do not appear to be real. 
    
\section{Summary and conclusion}

In Section~2 we pointed out that neither of the three recent studies 
involving NGC\,6994 has provided truly convincing arguments for accepting 
or rejecting the hypothesis that it is a physical stellar system. 
In Section~3, we first showed that NGC\,6994 is apparent as a stellar 
ensemble only by the concentration of a few stars brighter than 12.5\,mag.
A corresponding significant overdensity of fainter stars does not exist.
We then deduced from the number density of bright stars around 
NGC\,6994 that the relative frequency of close quadruples of 
stars like that of NGC\,6994 in a random distribution of field stars is 
low, but not low enough to render the random occurence of such a quadrupel 
impossible. 
In Section 4 we demonstrated that the proper motion data from the Tycho-2 
catalog provide a clear indication against the physical nature of the 
ensemble. 
The proper motions do not show any characteristic concentration and instead 
point towards a large spread in the tangential space velocities of the stars
that is consistent with the motions of the surrounding field stars. 
In Section 5 we presented and discussed radial velocities, atmospheric 
parameters and absolute magnitudes derived from high-resolution spectra
of the six obvious bright member candidates of NGC\,6994. We find that 
the sample contains three dwarfs, two giants and one subgiant. The radial 
velocities and absolute magnitudes yield further strong evidence against 
the hypothesis of a system of physically related stars. 
First, the radial velocities of the stars are substantially different from 
each other, which is unlikely to be due to binary orbital motion since 
reobservation at a second epoch assures that there are no significant 
variations with time. Second, the distance moduli of the stars which result 
from their spectroscopically determined absolute magnitudes and their 
apparent brightness do not agree to each other. The latter shows that the 
stars are not clustered at a certain location along the line of sight.  
Taking together the results from proper motions and spectroscopy we 
conclude that neither the sample as a whole nor 
a subgroup in it forms a physical stellar system. 
Instead the asterism called NGC\,6994 appears to be a rare occurrence of a 
random grouping of a few physically unrelated stars in projection on the 
celestial sphere.
This conclusion confirms the conjecture of Carraro (2000) who had doubted 
the existence of a cluster or cluster remnant because of the lack of a 
characteristic sequence in the color-magnitude diagram.\\

\begin{acknowledgements} We thank the Observatoire de Haute Provence 
and the GDR ``Galaxies'' for allocating observing time to our various
open cluster projects. M.O. thanks Eva Grebel for reading and commenting 
on the manuscript. 
\end{acknowledgements}

  \begin{table*}
 \vspace{2cm}
 \caption[]{Cluster member candidates in NGC\,6994:
Photometry, positions and proper \\ \hspace*{1.4cm} motions from the Tycho-2 catalog}
  \label{tab1}
  \begin{tabular}{cccccrr}
  \hline
  \noalign{\smallskip}
   TYC & $V_T$ & $(B-V)_T$ &$\alpha_{ICRS}$ & $\delta_{ICRS}$ & 
$\mu_\alpha \cos\delta$ & $\mu_\delta$ \\ 
   5778-&\multicolumn{2}{c}{[mag]}& [ $h\ m\ s$ ]& [ $^\circ$ $'$ $''$ ]&\multicolumn{2}{c}{[\masyr]} \\ 
  \noalign{\smallskip}
  \hline
  \noalign{\smallskip}
 0802 & 10.48 & 1.04 & 20 58 56.77 & $-$12 38 30.4 & $ +3.2\pm 1.5$ & $-17.2\pm 1.5$ \\ 
 0509 & 11.32 & 0.54 & 20 58 57.59 & $-$12 37 45.8 & $+15.4\pm 1.7$ & $ -8.0\pm 1.8$ \\ 
 0594 & 11.90 & 1.94 & 20 58 53.32 & $-$12 37 54.5 & $ +1.3\pm 3.3$ & $ -6.1\pm 3.4$ \\ 
 0492 & 11.94 & 0.76 & 20 58 54.79 & $-$12 38 04.2 & $ -1.9\pm 2.1$ & $-10.8\pm 2.2$ \\ 
 0565 & 12.36 & 0.38 & 20 58 59.68 & $-$12 41 43.7 & $+14.9\pm 2.0$ & $ +9.3\pm 2.1$ \\ 
 0796 & 11.16 & 1.27 & 20 58 52.50 & $-$12 42 46.5 & $ -7.6\pm 2.2$ & $ -5.0\pm 2.2$ \\ 
  \noalign{\smallskip}
  \hline
  \end{tabular}
\end{table*}

\vspace*{3cm}


  \begin{table*}
  \vspace{2cm}
  \caption[]{Results from high-resolution spectroscopy}
  \label{tab2}
  \begin{tabular}{ccccccccc}
  \hline
  \noalign{\smallskip}
  TYC & Date of& S/N & $V_r$ & $T_{eff}$ & $\log g$ & [Fe/H] & $M_V$&$(m-M)_V$\\
  5778-   & Observation &     & [km/s] & [K] & & & [mag]& [mag] \\ 
  \noalign{\smallskip}
  \hline
  \noalign{\smallskip}
0802  & 04/07/2000 &  8.2 & $-26.6$ & $4720\pm 60$ & $2.64\pm 0.05$ &$-0.22\pm 0.06$&$0.80\pm 0.11$&  9.58 \\
0802  & 10/08/2001 & 10.4 & $-26.9$ & $4740\pm 15$ & $2.57\pm 0.05$ &$-0.24\pm 0.07$&$0.91\pm 0.06$&  9.47 \\
  \noalign{\smallskip}
0509  & 04/07/2000 &  6.4 & $-53.8$ & $5887\pm 50$ & $4.07\pm 0.02$ &$+0.03\pm 0.03$&$3.65\pm 0.17$&  7.62 \\
0509  & 13/08/2001 & 16.4 & $-54.1$ & $5992\pm 41$ & $4.01\pm 0.05$ &$+0.01\pm 0.02$&$3.58\pm 0.14$&  7.69 \\
  \noalign{\smallskip}
0594  & 05/07/2000 &  8.6 & $-21.5$ & $4882\pm 64$ & $3.19\pm 0.11$ &$-0.17\pm 0.07$&$2.29\pm 0.37$ &  9.42 \\
0594  & 11/08/2001 &  7.0 & $-21.5$ & $5049\pm 44$ & $3.09\pm 0.13$ &$-0.08\pm 0.04$&$2.31\pm 0.45$ &  9.40 \\
  \noalign{\smallskip}
0492  & 04/07/2000 &  8.0 & $ -8.9$ & $5707\pm 68$ & $4.21\pm 0.12$ &$+0.01\pm 0.03$&$4.69\pm 0.14$ &  7.17 \\
0492  & 11/08/2001 & 11.0 & $ -8.9$ & $5762\pm 14$ & $4.32\pm 0.12$ &$-0.05\pm 0.05$&$4.70\pm 0.14$ &  7.16 \\
  \noalign{\smallskip}
0565  & 05/07/2000 &  5.4 & $-30.6$ & $6234\pm 25$ & $4.00\pm 0.02$ &$-0.20\pm 0.06$&$2.99\pm 0.14$ &  9.33 \\
0565  & 12/08/2001 & 13.5 & $-31.4$ & $6156\pm 42$ & $4.00\pm 0.02$ &$-0.36\pm 0.05$&$3.06\pm 0.15$ &  9.26 \\
  \noalign{\smallskip}
0796  & 05/07/2000 & 12.4 & $+34.7$ & $4748\pm 42$ & $2.60\pm 0.04$ &$-0.14\pm 0.04$&$0.87\pm 0.11$ & 10.17 \\
0796  & 10/08/2001 &  7.5 & $+34.8$ & $4770\pm 25$ & $2.56\pm 0.04$ &$-0.17\pm 0.05$&$0.82\pm 0.14$ & 10.22 \\
  \noalign{\smallskip}
  \hline
  \noalign{\smallskip}\noalign{Note: 
  S/N is the signal to noise ratio at 555 nm}
  \end{tabular}
  \end{table*}

\clearpage


 \begin{figure}[t]
 \centering
 \includegraphics[bb=65 210 470 590, width=8.8cm,clip=true]{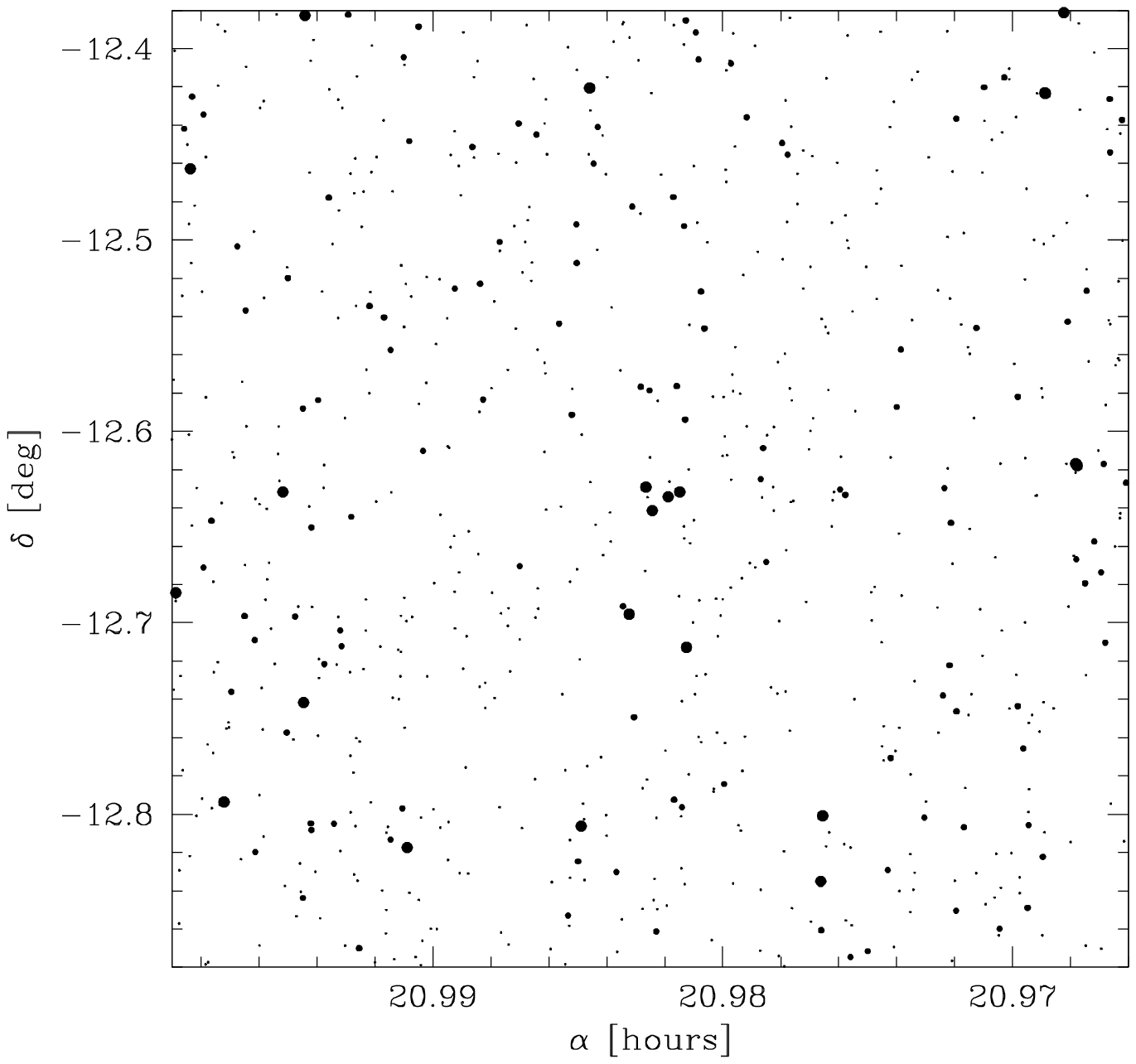}
 \caption{Distribution of stars from the USNO-A2.0 catalogue 
 (POSS-I) in a $30' \times 30'$ around NGC\,6994. The dot 
 sizes indicate magnitudes in three ranges: $R\le\ 12.5$ (fat), 
 $12.5 < R \le\ 15.0$ (medium), $15.0 < R \le\ 17.5$ (small). 
  }
 \label{fig1}

 \vspace*{.5cm}

 \centering
 \includegraphics[bb=65 185 470 645, width=8.8cm,clip=true]{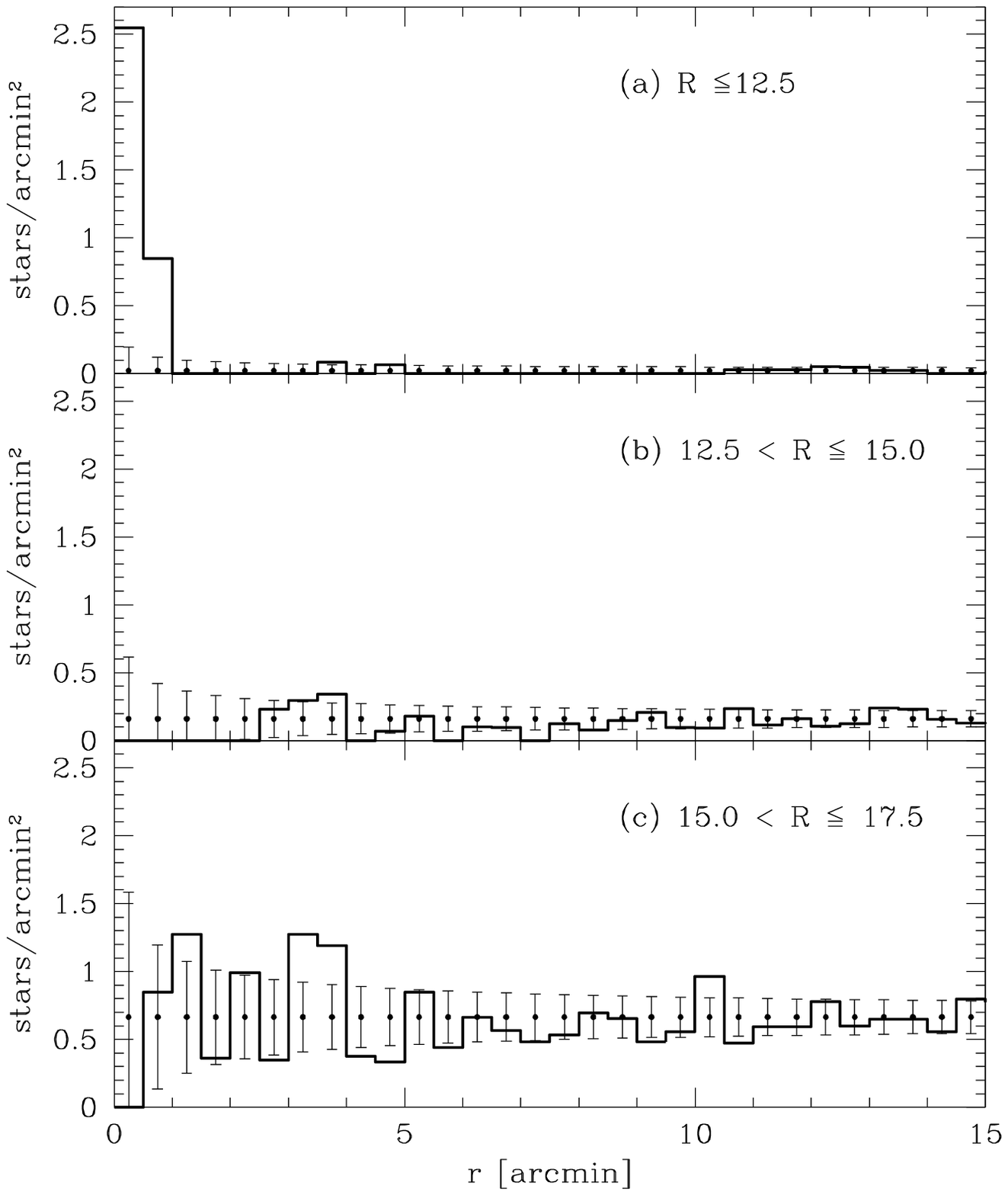}
 \caption{Histogram of the surface density of the stars shown in Fig.1, 
 measured by counts in $0\farcm 5$ wide annuli centered on 
 $\alpha = 20^h58^m55^s$, $\delta = -12^\circ 38' 04''$. 
 Dots with error bars indicate the mean surface density of stars 
 in the surrounding field and the expected $1\sigma$ range of statistical 
 fluctuations around this mean for each distance bin. 
 Only stars with $R \le 12.5$ show an enhanced surface density near the 
 position of NGC\,6994. The histograms for the fainter stars are 
 - within the statistical uncertainties - consistent 
 with a constant surface density that is equal to the density of field 
 stars. 
 \label{fig2}}
 \end{figure}



 \begin{figure}[t]
 \centering
 \vspace*{5cm}
 \includegraphics[width=8.8cm,clip=true]{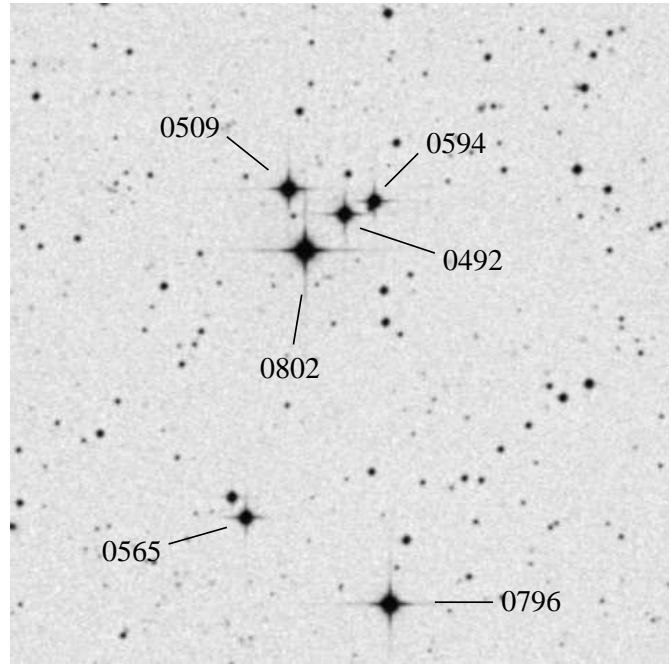}
 \caption{Identification map ($8' \times 8'$, image from DSS-2) for 
 the member candidates of NGC\,6994 investigated in this paper.
 North is up and east to the left. 
 The labels give the star numbers assigned in the Tycho-2 catalog 
 (in the sense TYC 5778-$<$...$>$). Data for these stars are given 
 in Tables 1 and 2.}
 \label{fig3}
 \end{figure}


\clearpage


 \begin{figure*}[t]
 \centering
 \includegraphics[bb=30 150 570 700,width=18cm,clip=true]{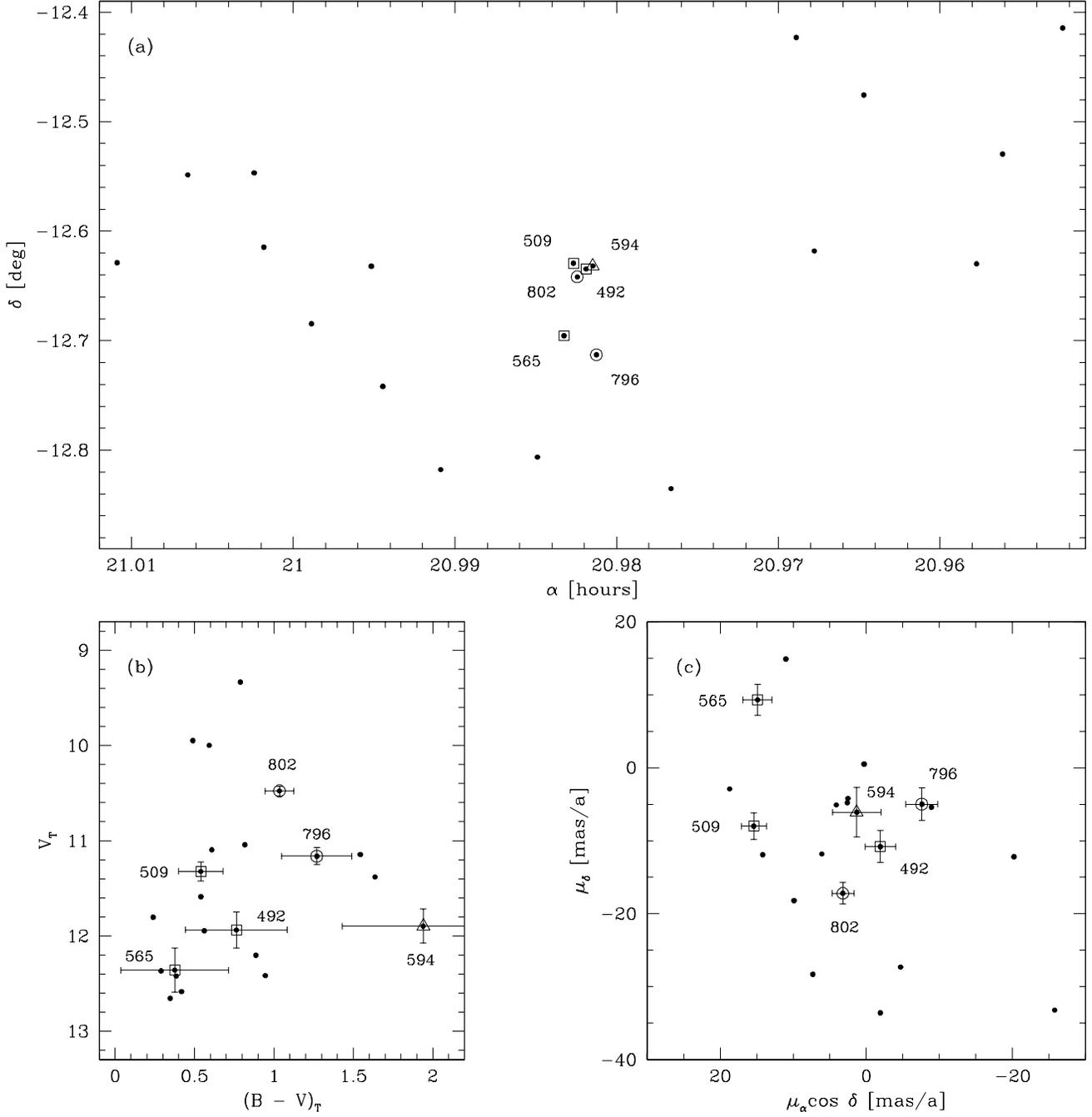}
 \caption{Position, photometry and proper motions of all Tycho-2 stars 
(H{\o}g et al.\ 2000) in a $54'\times 30'$ field centered on NGC 6694. 
(a) Distribution on the sky. (b) Color-magnitude diagram. (c) Vector 
point plot of Tycho-2 proper motions and proper motion errors. 
In each panel the spectroscopically observed stars are marked by open 
symbols and labelled with their TYC number (see Tables 1 and 2). The 
different symbol types distinguish dwarfs (open squares), 
giants (open circles), and one subgiant (triangle). }
 \label{fig4}
 \end{figure*}

%

\end{document}